\documentclass[a4paper,14pt]{article}

\usepackage[utf8]{inputenc}   
\usepackage[english]{babel}

\usepackage{graphicx} 
\usepackage{multirow}
\usepackage{array} 

\usepackage{hyperref} 

\usepackage[usenames,dvipsnames]{pstricks}

\begin{document}
\pagestyle{plain}
\pagenumbering{arabic}

\selectlanguage{english}

\title{The role of quantum-well states and carrier scattering times on discontinuities in opto-electrical characteristics of SCH lasers.}
\author{Zbigniew Koziol\footnote{zbigniew{@}ostu.ru}, Sergey I. Matyukhin\footnote{sim1{@}mail.ru},\\ 
State University - Education-Science-Production Complex, \\
29 Naugorskoye Shosse, Oryol, 302020, Russia,\\
and Evgeniya A. Buduleva \footnote{janesname{@}yandex.ru},\\
Oryol State University,\\
95 Komsomolskaya Street, Oryol, 302026, Russia}


\maketitle

\baselineskip=4.00ex 

\begin{abstract}
Drift-diffusion computer simulations model available in Synopsys' Sentaurus TCAD 
is used to study electrical, I-V, and optical, I-L, 
characteristics of separate-confinement heterostructure laser based 
on $AlGaAs$. We investigate the role of the width and depth of Quantum Well (QW) 
active region, below and above the lasing threshold. 
The device properties depend on both, the number of bound QW states and on closeness 
of the highest bound states to conduction or valence band offset.
The lasing action may not exist at certain widths or hights of QW, 
and the threshold current is a discontinuous function of these parameters.
The effects are more pronounced at low temperatures. Discontinuities in characteristics
are found, at certain conditions, in temperature dependencies as well. 
The carriers scattering time on QW is shown to have the crucial role 
on amplitude of discontinuities. 
\end{abstract}

 
\maketitle


\section{Introduction}
\label{sec:intro} 

Computer modelling of electronic devices is a relatively 
new approach towards study of physical phenomena occurring there as well
optimizing their technical characteristics. Methodologicaly, this field of
scientific and engineering activity may be placed between theory and experiment,
not belonging however to either of them:
persuing research of that kind requires theoretical understanding of physics
of microscopic processes and may work as a helpfull tool in interptretation 
of experimental data. Contrary to sometime met thinking, computer modelling
can not replace theory or experiment. In some situations, results
of that research may provide an inspiration for understanding or testing physical phenomena: it is
easier, faster and less expensive to perform modelling than experiments, and
we are not restricted that much by, often large, inaccuracy of experimental data that may hide
insightfull details. 

When performing modelling $AlGaAs$ SCH lasers with Synopsys' Sentaurus TCAD \cite{tcad}, we noticed unexpected steps 
in some of their characteristics (threshold current $I_{th}$ versus the width of quantum well $d_a$ (\cite{Matukhin}),
or versus it's height, etc). Analyses of results led us to conclusion that observed discontinuities 
occur when the most upper bound QW state crosses the conduction 
or valance band offset energy. Following that idea, we guessed that the effects may manifest
itself in temperature dependence of other physical quantities as well, if laser parameters are choosen for that
properly. In this work we show that the discontinuities are found also below the lasing threshold current,
in their $I-V$ characteristics or in gain (or loss) versus current. This article is 
continuation of our efford to understand better the nature of these physical phenomena (\cite{arxive}). 
Here we show results of computer modelling of the role of quantum well (QW) scattering times.

\section{Modelling}
\label{sec:modelling} 

The laser we are modelling has dimensions, structure and doping 
as described by Andreev, et al. \cite{Andrejev_1}, \cite{Andrejev_2}.
The lasing wavelength is $808 nm$, the lasing offset voltage $U_0$ is $1.56-1.60 V$, 
differential resistance just above the lasing offset, $r=dU/dI$, is $50-80 m\Omega$, 
threshold current $I_{th}$ is $200-300 mA$,
slope of optical power, $S=dL/dI$, is $1.15-1.25 W/A$, and left and right mirror reflection 
coefficients $R_l$ and $R_r$ are $0.05$ and $0.95$. The reference laser has the width of QW, $d_a$, 
of $12 nm$ and both waveguides' width is $0.2 \mu m$. 
In order to reproduce laser characteristics in computational results, we
played with several variables available in Synopsys. The critical one is $A_{ph}$ 
- the effective surface area factor in $Physics$ section of Synopsys command file. 
An agreement between experiment and calculation is reached for $A_{ph}$ of about $0.059$. 
This low value of $A_{ph}$ should not be surprising:  
the physics we are dealing with here, in particular in waveguide and in QW regions, 
depends on ballistic transport as well, 
while the computational model we use is derived from drift-diffusion equations, 
modified for dealing with transport through QW as discussed in the next section.

\begin{table}[h]
\begin{center}
\begin{tabular}{c|c|c|c}
\hline
\#  		&	C [$cm^3/s$] 	&  $\alpha_n$ [$cm^{-2}$]	&  $\alpha_p$ [$cm^{-2}$]\\
\hline
A 			&	$2.0 \cdot 10^{-10}$	&	$1 \cdot 10^{-18}$		&	$2 \cdot 10^{-18}$	\\
B 			&	$2.0 \cdot 10^{-10}$	&	$5 \cdot 10^{-17}$		&	$1 \cdot 10^{-18}$	\\
C 			&	$2.0 \cdot 10^{-10}$	&	$1.5 \cdot 10^{-18}$	&	$3 \cdot 10^{-18}$	\\
D 			&	$1.0 \cdot 10^{-10}$	&	$1.5 \cdot 10^{-18}$	&	$3 \cdot 10^{-18}$	\\
E 			&	0 					&	$1.5 \cdot 10^{-18}$	&	$3 \cdot 10^{-18}$	\\
F 			&	0					&	$1.5 \cdot 10^{-18}$	&	$3 \cdot 10^{-18}$	\\
\hline
\end{tabular}
\caption{A few sets of simulation conditions ($A-F$) 
for data shown in Figures \ref{qw300_talk0A} and \ref{qw300_talk1}. 
$C$ is radiative recombination rate (Eq. \ref{Radiative_recombination_rate}).
$\alpha_n$, and $\alpha_p$ are coefficients of free carrier absorption formula (\ref{Free_carrier_absorption}).
Temperature for all cases is $T=300 K$, electron and hole scattering times are assumed 
$8\cdot 10^{-13} s$ and $4\cdot 10^{-13} s$, respectively, and electron and hole mobility 
$9200 cm^2/Vs$ and $400 cm^2/Vs$, respectively. No additional light scattering mechanisms are considered.  
}
\label{table_4}
\end{center}
\end{table}

\begin{figure}
\begin{center}
\includegraphics[width=0.85\columnwidth]{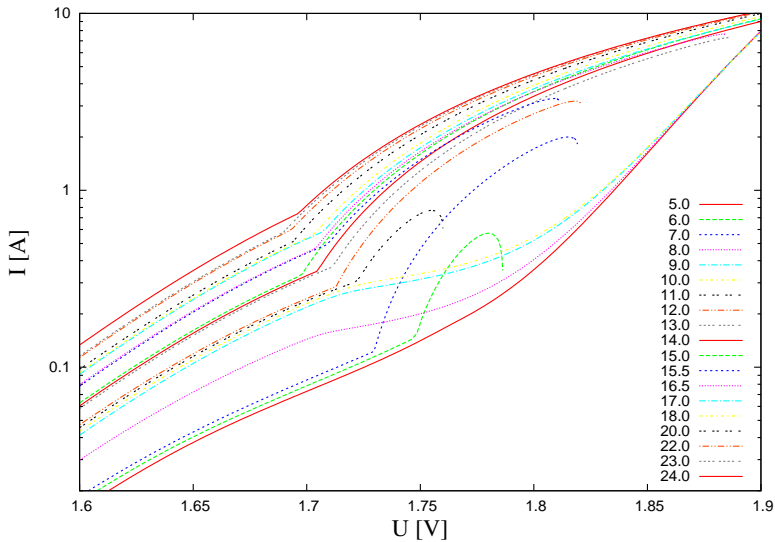}
      \caption{
	Typical $I-V$ characteristics computed at $T=300$. $OpticalLoss$ parameter is assumed $0$,
	no radiative recombination, free carrier scattering rate parameters $\tau_n$ and $\tau_p$
	are $8 \cdot 10^{-13} s^{-1}$and  $4\cdot 10^{-13}  s^{-1}$, with electron and hole mobolities
	$9200 cm^2/Vs$ and $400 cm^2/Vs$. 
	The legend describes width of QW (in $nm$), from $5 nm$ from right-bottom curve to 
	$24 nm$ for uppermost curve.
}
	 \label{qw300_000}
\end{center}
\end{figure}

Other parameters available in Synopsys, important in this case, are these related 
to light absorption and carrier scattering. Experiment shows that absorption coefficient 
is of the order of $1 cm^{-1}$ (\cite{Andrejev_1}). It is argued that in AlGaAs lasers 
the main contribution to light absorption is due to photon scattering on free carriers, with 
the free carrier absorption coefficient, $\alpha_{fc}$, given by:

\begin{equation}\label{Free_carrier_absorption}
	\alpha_{fc} = \left(\alpha_n \cdot n + \alpha_p \cdot p\right) \cdot L,
\end{equation}

where $n$ and $p$ are the electron and hole density, and $L$ is light intensity. We choose 
in our calculations such values of $\alpha_n$ and 
$\alpha_p$ that an effective absorption coefficient obtained would be close to that experimental one.

It is important also to have a reasonable value of radiative recombination rate, 
$R_r$, which is assumed to be described by: 

\begin{equation}\label{Radiative_recombination_rate}
	R_{r} = C \cdot \left(n \cdot p - n_{i_{eff}}^2\right),
\end{equation}

where $n_{i_{eff}}$ describe the effective intrinsic density, and $C$ is a parameter available for changes.

Typical $I-V$ characteristics computed at $T=300$ are shown in Figure \ref{qw300_000}, 
for a broad range of QW widths. For current near the lasing threshold current $I_{th}$ 
(i.e. for voltage near the lasing offset voltage $U_0$), which correspond to a kink in $I-V$, 
for most of these curves the results are very well approximated by a phenomenological modified 
exponential relation (\cite{Koziol}):

\begin{equation}\label{exponential_6_parameters}
\begin{array}{ll}
        I(U) = I_{th} \cdot exp(A\cdot (U-U_0) + B \cdot (U-U_0)^2),~~~ for ~ U< U_0\\
        I(U) = I_{th} \cdot exp(C\cdot (U-U_0) + D \cdot (U-U_0)^2),~~~ for ~ U> U_0
\end{array}
\end{equation}

where $I_{th}$, $U_0$, as well $A$, $B$, $C$, and $D$ are certain fiting parameters.

\begin{figure}
\begin{center}
\includegraphics[width=0.85\columnwidth]{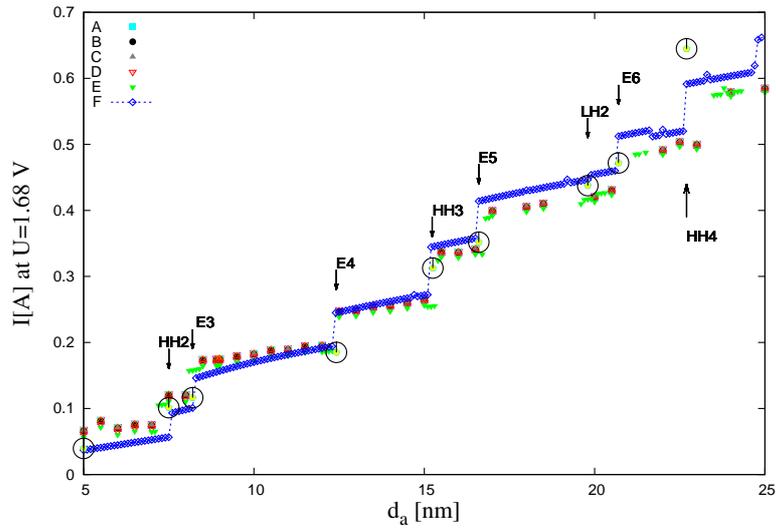}
      \caption{
	Current as a function of QW width derived at constant voltage from datacurves 
	as these shown in Figure \ref{qw300_000} and described in Table \ref{table_4}. 
	The solid line for datapoints $F$ is to guide the eyes, only. $F$ is computed for constant 
	$Al$ concentration in QW of $8\%$, while all other datasets ($A-E$) are computed with
	such a concentration of $Al$ in QW that lasing wavelength will remain constant ($808 nm$)
	when QW width changes. The arrows are for datapoints $F$.
	From results of a separate analysis, not shown in this Figure, it follows
	that the relative hight of steps and position of these steps does not depend on voltage at which current is messured, 
	even though current values may change as much as 100 times.
}
\label{qw300_talk0A}
\end{center}
\end{figure}

There straightforward interpretation of these curves around the lasing threshold, 
since a strong, nonlinear interplay between the effects of carrier transport and scattering 
takes place, with light absorption as well. However, we may notice an interesting feature 
for parts of curves below the lasing threshold. While
the width of QW, $d_a$, changes (nearly) monotonically, the curves however are grouped 
into a few sets such that they nearly coincide together, within each group.

\begin{figure}
\begin{center}
\includegraphics[width=0.85\columnwidth]{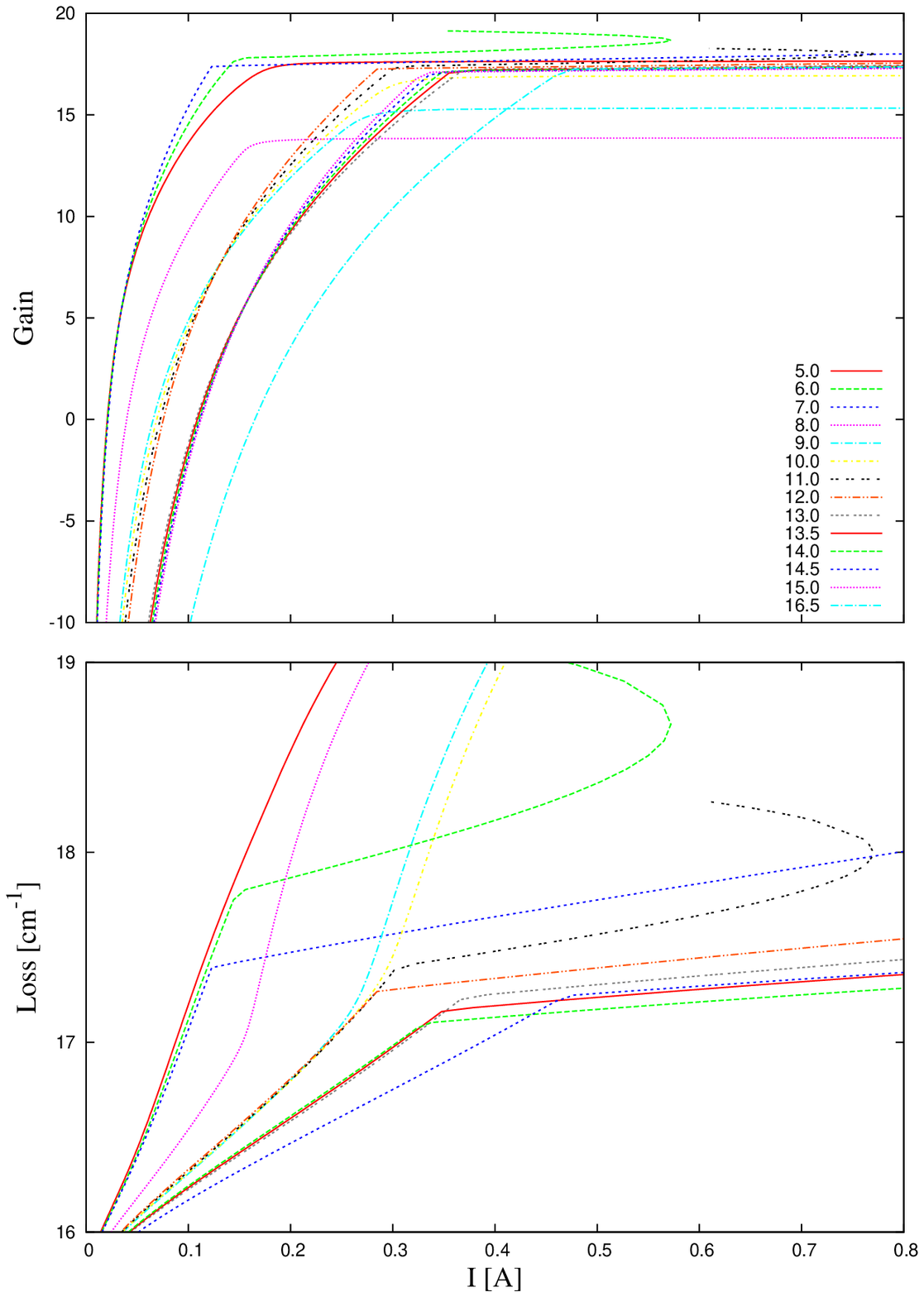}
      \caption{
	Gain (upper figure) and loss computed in similar conditions as the data in Figure \ref{qw300_000}.
	The legend describes width of QW (in $nm$).
}
\label{qw300_014A}
\end{center}
\end{figure}

A very similar feature is observed when gain or loss is drawn as a function of current, for many
widths of active region (Figure \ref{qw300_014A}). Again, we see that datacurves for current
below the lasing threshold (that corresponds to kinks in curves) 
are grouped into a few sets such that they nearly coincide together within each group.
If gain or loss were drawn as a function of voltage, however, we would not see such a grouping.

Therefore, we conclude that below the lasing threshold, 
current as a function of QW width at constant voltage derived from data like these in Fig. \ref{qw300_000},
or gain or loss as a function of QW width, at constant current values, also below the lasing threshold,
will follow step-like functions.

\begin{figure}
\begin{center}
\includegraphics[width=0.85\columnwidth]{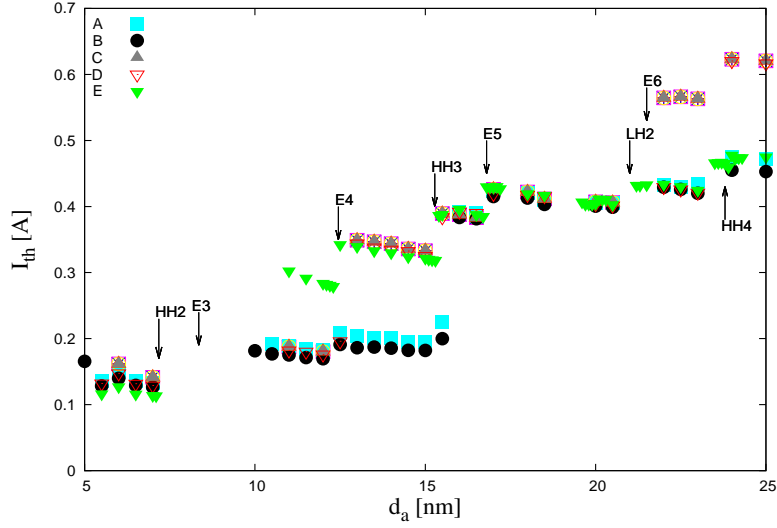}
      \caption{
	Lasing threshold current as a function of QW width for datasets 
	as these shown in Figure 
\ref{qw300_000} and described in Table 
\ref{table_4}. 
	The arrows are at positions close to but not identical to these in Figure 
\ref{qw300_talk0A}.
}
\label{qw300_talk1}
\end{center}
\end{figure}

\begin{figure}
\begin{center}
\includegraphics[width=0.85\columnwidth]{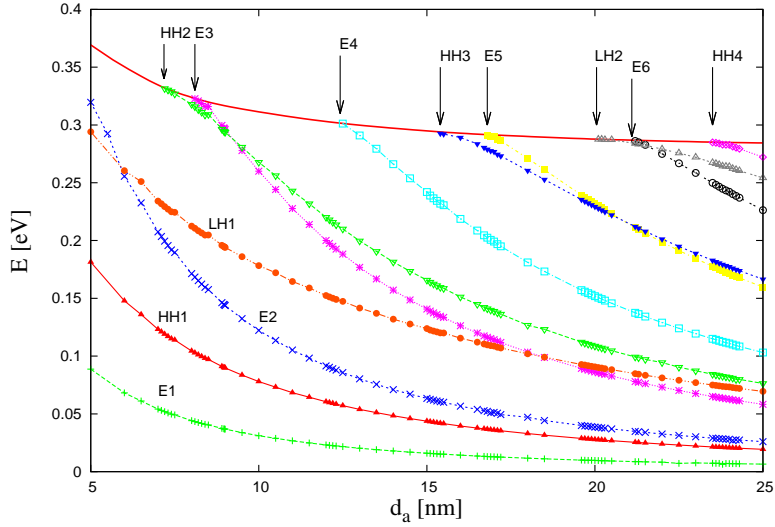}
      \caption{
T=300K. Conduction and valence band offset energy, ($E_{CBO}$ and $E_{VBO}$), 
and electron ($E_n$), light- and heavy hole energies ($LH_n$ and $HH_n$) in QW, 
as a function of QW width. Hole energies and $E_{VBO}$ have been scaled up 
by a factor 28 to obtain coincidnce with electron energy scale (i.e., $E_{CBO}$ and $E_{VBO}$
curves are the same in this Figure). 
}\label{qw300_energy0}
\end{center}
\end{figure}

This is illustrated in Fig. \ref{qw300_talk0A}, where current as a function of QW width derived 
at constant voltage from datacurves similar to these as in Figure \ref{qw300_000} is shown. 
Figure \ref{qw300_talk0A} presents data computed at different conditions, and marked from $A$ to $F$, 
for several combinations of free carrier scattering coefficients, $\alpha_n$ and $\alpha_p$,
and values of $C$, the radiative recombination parameter, as described in Table \ref{table_4}.
Moreover, dataset $F$ differs from datasets $A$-$E$. The last are computed assuming changing 
Al concentration in waveguides (when Al in QW is kept constant) in such a way that the lasing 
wavelength does not change with the change of QW width. The dataset $F$ is computed for constant 
Al concentration in waveguides of $33 \%$. The solid line in Fig. \ref{qw300_talk0A} is drawn through
datapoints $F$, and arrows there refer to curve $F$ as well, and indicate positions of bound QW energy states crossing
the conduction- or valance band offset energies. Positions of bound QW energy states for cases $A$-$E$,
as illustrated in Fig. \ref{qw300_energy0}, are very close to but not identical. The results of Figure 
\ref{qw300_energy0}, described in more details earlier (\cite{Koziol2}), were found also 
to be exactly the same as computed by using nextnano software (\cite{Birner}).

\begin{figure}
\begin{center}
\includegraphics[width=0.85\columnwidth]{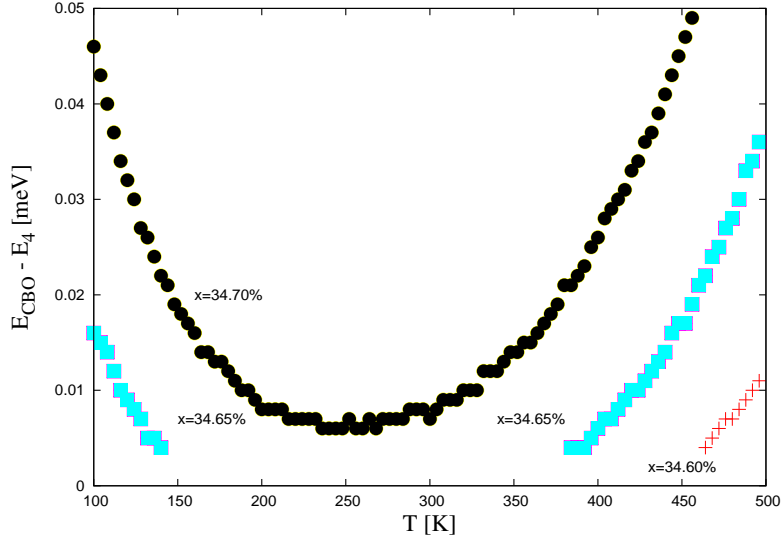}
      \caption{The energy difference between conduction band offset energy $E_{CBO}$
and the closest to it bound electron state in QW, $E_4$, as a function of temperature, 
when active region concentration is 8\% of Al, for three values of waveguide Al concentraion:
$34.60\%$, $34.65 \%$, and $34.70\%$.
}
\label{temp-swap00}
\end{center}
\end{figure}

The step-like features are preserved also in $I_{th}(d_a)$ dependencies, 
as illustrated in Fig. 
\ref{qw300_talk1}. There, however, effects of carrier scattering and light 
absorption smear-out the picture. It is useful to notice that at some values of QW width no lasing 
action is reached, and therefore the datapoints in that Figure are not available for all QW widths.

Changes of QW height (caused by difference of Al concentration in QW and waveguides) cause very 
similar step-like dependencies. Moreover, the effects are in some situations more clear and pronounced
at low temperatures. We did modelling for T=77.6 K to confirm their existence.

\begin{figure}
\begin{center}
\includegraphics[width=0.85\columnwidth]{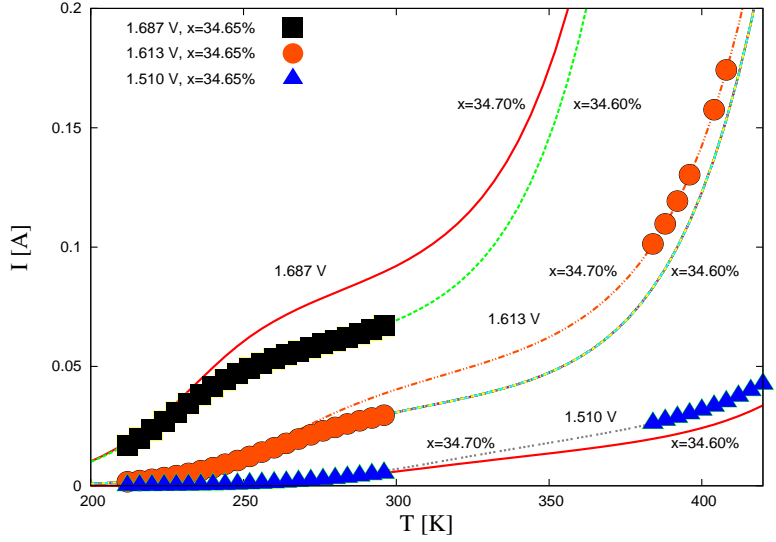}
      \caption{Temperature dependence of current for three values of voltage applied 
(1.510V, 1.613V, and 1.687V; for each of these this is below the lasing threshold),
and for free values of $Al$ concentrations in waveguide, the same as in Figure 
\ref{temp-swap00}.
}
\label{temp-swap06}
\end{center}
\end{figure}

Moreover, with a careful design of laser structure (content of Al in QW and waveguides) it is possible to 
find the evidence of the effect in temperature dependence of current, when measurements are performed
at constant voltage. We found such Al concentrations when
the number of QW bound states changes with temperature swap. Figure 
\ref{temp-swap00} shows how the uppermost 
bound electron state energy, $E_4$ in this case, differs from $E_{CBO}$ 
(energy levels of other QW bound states do not play a significant role in this case), 
for a three Al concentrations in waveguide,
when Al concentration in QW is $8\%$. For Al concentration $34.70 \%$, the $E_4$ energy level exists always 
through temperature swap studied. For Al concentration $34.65 \%$, the $E_4$ energy level does not exist 
between around $150$ and $370 K$. For Al concentration $34.60 \%$, it exists at temperatures higher than
about $470 K$, only. This has profound implications on $I(T)$ dependencies measured at constant voltage 
for these three different Al concentrations of Al in waveguide, as Figure 
\ref{temp-swap06} illustrates.
For $34.60 \%$ and $34.70 \%$ of Al content, we observe continuous $I(T)$ curves. However, for 
$34.65 \%$ of Al content, at low temperatures $I(T)$ results fall on curve that has been computed for $34.60 \%$
of Al, and at high temperatures they fall on the curve computed for $34.70 \%$ of Al.

\section{The role of QW scattering times.}
\label{sec:scattering} 

The modeling results reported above were computed by using default (in Synopsys TCAD) values of parameters 
related to carriers scattering times on QW (i.e. time of living of carriers in bound QW states).
An approach used in Synopsys is to have constant vaues of these parameters (which however may be adjusted by user).

Existing theoretical and other computational results indicate that carriers scattering times depend on QW width
(\cite{Ferreira}, \cite{Birner2}). Scattering on longitudinal optical phonons, LO, is considered as the main mechanism.
Birner \cite{Birner2}) computed lifetime of electron states for transitions between the initial state $E_2$
and the final ground state $E_1$), i.e. for an intersubband transition, for different quantum well widths at $T=0$.

The $nextnano^3$ calculations of Birner (\cite{Birner2}) are in good agreement with these of Ferreira and G. Bastard (\cite{Ferreira}).
For quantum well widths smaller than about 5.4 nm, only the ground state is confined and $E_2$ is unbound.
For quantum well widths larger than about $18 nm$, the transition energy becomes smaller 
than the LO phonon energy and scattering through the emission of an LO phonon is not possible any more.

In order to have a better view on what is the effect of scattering times on the amplitude of steps 
(and their existence) in opto-electrical characteristics of the device, we performed simulations for several sets of values 
of parameters available in Synospsys, as described in Table \ref{table_5}.

\begin{table}[h]
\begin{center}
\begin{tabular}{c|c|c|c|c}
\hline
\#  		&	QWeScatTime  	&  QWhScatTime 	&  eQWMobility &  eQWMobility\\
		&	[$s$]		&	[$s$]		&	[$cm^2/Vs$]&	[$cm^2/Vs$] \\
\hline
a 			&	$8.0 \cdot 10^{-13}$	&	$4 \cdot 10^{-13}$		&	$9200$	&	$400$	\\
b 			&	$2.0 \cdot 10^{-13}$	&	$2 \cdot 10^{-13}$		&	$9200$	&	$400$	\\
c 			&	$1.0 \cdot 10^{-13}$	&	$5 \cdot 10^{-14}$		&	$9200$	&	$400$	\\
d 			&	$2.0 \cdot 10^{-12}$	&	$2 \cdot 10^{-12}$		&	$9200$	&	$400$	\\
e 			&	$4.0 \cdot 10^{-12}$	&	$2 \cdot 10^{-12}$		&	$9200$	&	$400$	\\
f 			&	$8.0 \cdot 10^{-13}$	&	$4 \cdot 10^{-13}$		&	$5000$	&	$200$	\\
g 			&	$8.0 \cdot 10^{-13}$	&	$4 \cdot 10^{-13}$		&	$1000$	&	$40$	\\
h 			&	$8.0 \cdot 10^{-13}$	&	$4 \cdot 10^{-13}$		&	$20000$	&	$800$	\\
\hline
\end{tabular}
\caption{A few sets of simulation conditions ($a-h$ in the first row) 
for data shown in Figures \ref{qw300_A_0} - \ref{qw300_A_3}. QW width is 12 nm.
}
\label{table_5}
\end{center}
\end{table}

\begin{figure}
\begin{center}
\includegraphics[width=0.85\columnwidth]{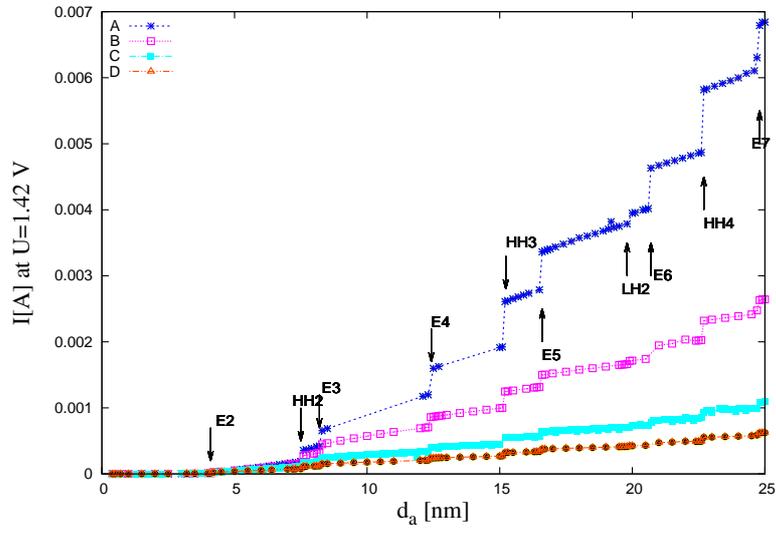}
      \caption{Height of steps does not depend in a noticable manner on parameters eQWMobility and hQWMobility.
      The data in this figure are all computed by using the same values of these parameters, 9200 and 400 $cm^2/Vs$, respectively.
	 The effect of QW scattering times is however pronounced. The data, for curves A to D (from top to bottom) were computed
	 with the following sets of (QWeScatTime, QWhScatTime): ($10^{-13}$, $5\cdot 10^{-14}$), ($8\cdot 10^{-13}$, $4\cdot 10^{-13}$), ($4\cdot 10^{-12}$, $2\cdot 10^{-12}$), and ($10^{-11}$, $5\cdot 10^{-12}$).
      where time is in seconds.
}
\label{qw300_BCDE}
\end{center}
\end{figure}

Typical results are shown in Figure \ref{qw300_BCDE}. 
Shorter QW scattering times lead to strong increase of current and pronounced steps 
in current as a function of QW width. It was found that the value of parameters 
eQWMobility and hQWMobility has no a noticable effect to these results.


\begin{figure}
\begin{center}
\includegraphics[width=0.85\columnwidth]{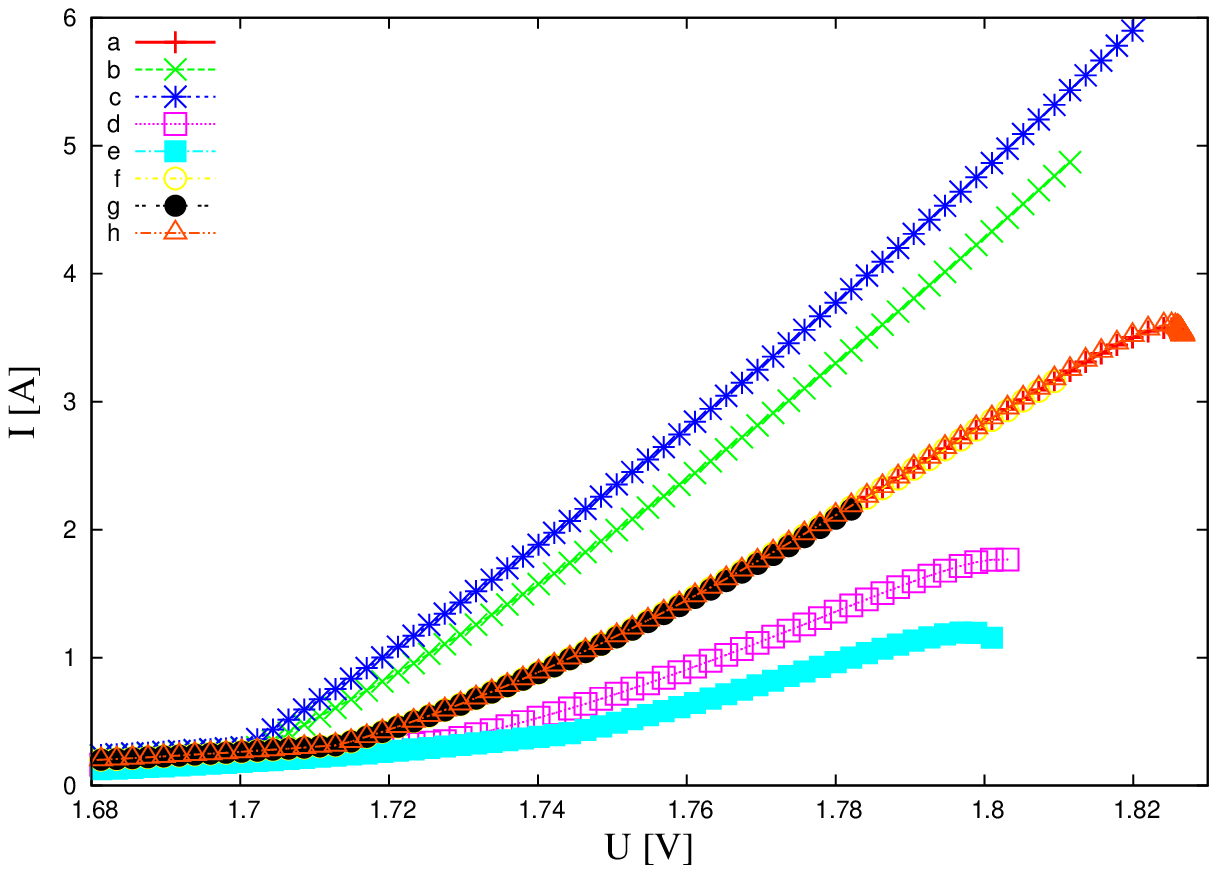}
      \caption{ I-V characteristics for a laser with 12 nm QW width, 
	with QWeScatTime, QWhScatTime, eQWMobility, and hQWMobility parameters, for curves from a to h, as described in
Table \ref{table_5}. The curves with the same values of QWeScatTime and QWhScatTime coincide together, regardles of 
of values of eQWMobility and hQWMobility.
}
\label{qw300_A_0}
\end{center}
\end{figure}

\begin{figure}
\begin{center}
\includegraphics[width=0.85\columnwidth]{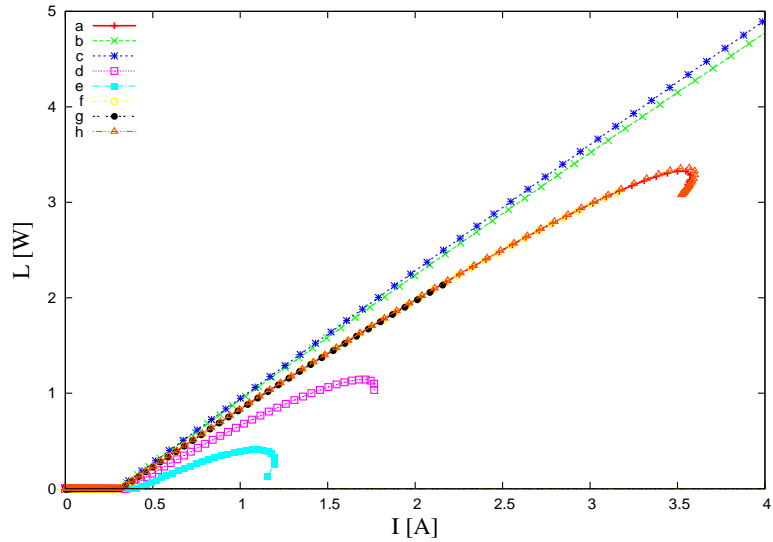}
     \caption{ L-I characteristics for a laser with 12 nm QW width, for cases a to h the same as these in Figure \ref{qw300_A_0}
	and described in Table \ref{table_5}
}
\label{qw300_A_2}
\end{center}
\end{figure}

\begin{figure}
\begin{center}
\includegraphics[width=0.85\columnwidth]{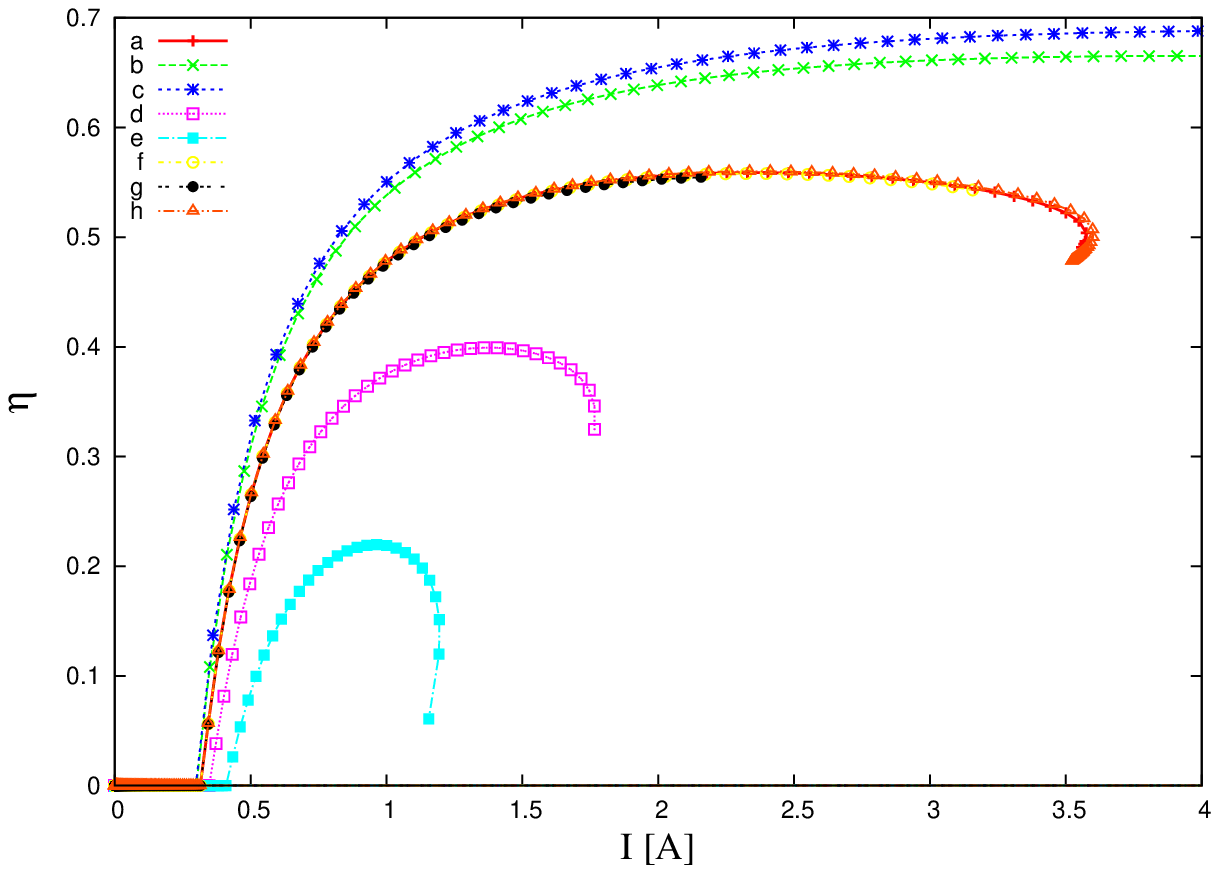}
     \caption{ Optical efficiency, $L/(U\cdot I)$ for a laser with 12 nm QW width, for cases a to h the same as 
	these in Figure \ref{qw300_A_0} and described in Table \ref{table_5}.
}
\label{qw300_A_3}
\end{center}
\end{figure}

For a completness of the results on modeling the role of QW scattering, we show also I-V characteristics (Figure \ref{qw300_A_0}),
lasing light intensity L-I (Figure \ref{qw300_A_2}) and optical efficiency curves (Figure \ref{qw300_A_3}) for a few sets of values of
QW scattering parameters.

\section{Discussion}
\label{sec:discussion} 

In case of tunneling energy barrier, transfer matrix approach is used 
to describe charge transport through it (\cite{Davies}, \cite{Piprek}). 
The interband tunneling current is written as

\begin{equation}\label{current_density}
\begin{array}{ll}
J \sim \int_{E_{min}}^{E_{max}} \cdot N(E) \cdot f(E) \cdot T(E) \cdot dE
\end{array}
\end{equation}

where $T(E)$ is energy-dependent tunneling rate, $N(E)$ is the density of states, 
$f(E)$ is the Fermi-Dirac distribution function, respectively,
and $E_{min}$ and $E_{max}$ are minimum and maximum carrier energies available.

\begin{figure}[h]
\begin{center}
\scalebox{0.9} 
{
\begin{pspicture}(0,-4.68)(11.561563,4.69)
\definecolor{color39}{rgb}{0.8666666,0.141176470,0.3647}
\definecolor{color39b}{rgb}{0.878431,0.07058823,0.27450}
\definecolor{color187b}{rgb}{0.08235,0.152941,0.996078}
\definecolor{color205b}{rgb}{0.00392,0.898039,0.3411764}
\definecolor{color0b}{rgb}{0.0823529,0.03529,0.19215686}
\definecolor{color0}{rgb}{0.45882352,0.0627450,0.917647}
\definecolor{color22}{rgb}{0.5529411,0.031372,0.376470588}
\psline[linewidth=0.04,linecolor=color0,fillcolor=color0b](4.0215626,2.32)(4.0215626,-3.58)(8.041562,-4.66)(8.061563,1.36)(8.061563,1.36)(8.061563,1.36)
\psline[linewidth=0.04cm,linecolor=color22](0.6015625,4.32)(4.0615625,2.26)
\psline[linewidth=0.04cm,linestyle=dotted,dotsep=0.16cm](1.6015625,-2.4)(9.801562,-2.46)
\psline[linewidth=0.06cm,linestyle=dashed,dash=0.16cm 0.16cm,arrowsize=0.05291667cm 2.0,arrowlength=1.4,arrowinset=0.4]{<->}(5.9415627,1.96)(5.9615626,-2.16)
\psdots[dotsize=0.4,linecolor=color39,fillstyle=solid,fillcolor=color39b,dotstyle=o](3.1215625,3.14)
\psarc[linewidth=0.06,linestyle=dashed,dash=0.16cm 0.16cm,arrowsize=0.05291667cm 2.0,arrowlength=1.4,arrowinset=0.4]{<->}(7.9915624,2.27){1.83}{0.0}{180.0}
\psarc[linewidth=0.06,linestyle=dashed,dash=0.16cm 0.16cm,arrowsize=0.05291667cm 2.0,arrowlength=1.4,arrowinset=0.4]{<->}(4.5615625,3.28){1.38}{0.0}{180.0}
\usefont{T1}{ptm}{m}{n}
\rput(2.00,-2.05){QW bound states}
\usefont{T1}{ptm}{m}{n}
\rput(1.5,1.4){QW continuum states}
\psline[linewidth=0.02cm,arrowsize=0.05291667cm 2.0,arrowlength=1.4,arrowinset=0.4]{<-}(5.9615626,2.62)(3.8615625,2.0)
\psdots[dotsize=0.4,linecolor=color39,fillstyle=solid,fillcolor=color39b,dotstyle=o](9.821563,2.08)
\psdots[dotsize=0.4,linecolor=color39,fillstyle=solid,fillcolor=color187b,dotstyle=o](5.9815626,-2.46)
\psdots[dotsize=0.4,linecolor=color39,fillstyle=solid,fillcolor=color205b,dotstyle=o](6.1215625,2.1)
\psdots[dotsize=0.4,linecolor=color39,fillstyle=solid,fillcolor=color205b,dotstyle=o](5.9415627,3.04)
\psline[linewidth=0.04cm,linecolor=color22](8.081562,1.3)(11.541562,-0.76)
\usefont{T1}{ptm}{m}{n}
\rput(7.8253126,3.6){Thermionic emission}
\usefont{T1}{ptm}{m}{n}
\rput(5.5,-0.21){Carrier Scattering}
\end{pspicture}
}
\caption{The model of carrier scattering at the quantum well used in Sentaurus (based on \cite{tcad}).}
\label{QWscattering}
\end{center}
\end{figure}
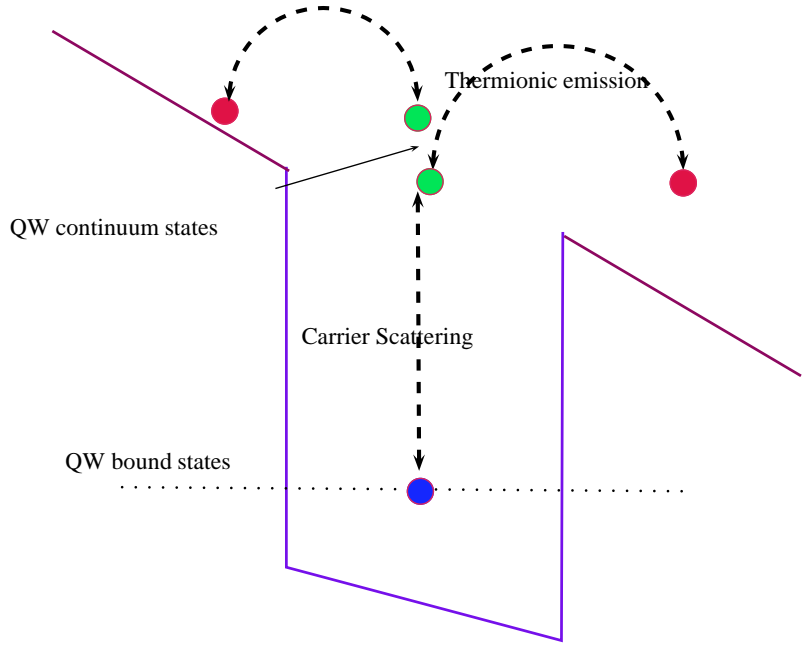

In Sentaurus, s simplified intuitive model is used to handle the physics of carrier scattering 
at the quantum well (Figure 
\ref{QWscattering}). The carrier populations are separated into bound and continuum states, 
and separate continuity equations are applied to both populations. The QW scattering model accounts for
the net capture rate, that is, not all of the carriers will be scattered into the bound states of the
quantum well. The electron capture rate from the continuum (subscript $b$) to the bound (subscript $b$) 
states is:

\begin{equation}\label{electron_capture}
\begin{array}{ll}
R = \int _{E_c}^{\infty} dE_c \int _{E_{b}}^{\infty} dE_b \cdot N_c(E_c) \cdot N_b (E_b) \cdot S(E_b,E_c) \cdot f_c (E_c) (1-f_b (E_b))
\end{array}
\end{equation}

where $E_c$ and $E_b$ is energy of lowest conduction band-, and bound QW electron states, 
$N(E)$ is the density-of-states, $S(E_b, E_c)$ is the scattering probability, and $f(E)$ is the
Fermi–Dirac distribution.
The reverse process gives the electron emission rate from the bound to continuum states:

\begin{equation}\label{electron_emission}
\begin{array}{ll}
M = \int _{E_c}^{\infty} dE_c \int _{E_{b}}^{\infty} dE_b \cdot N_c(E_c) \cdot N_b (E_b) \cdot S(E_b,E_c) \cdot f_b (E_b) (1-f_c (E_c))
\end{array}
\end{equation}

The net capture rate is $C = R - M$, and for very deep quantum wells (keyword $QWDeep$
must be used for that in Sentaurus) is known to be given by approximation:

\begin{equation}\label{capture_rate_deep}
C = R - M = \left(1 - exp(\eta_b - \eta_c)\right) \cdot \frac {n_c}{\tau}
\end{equation}

where $\eta_b = (-q \Phi_b - E_c)/k_B T$ and $\eta_c = ( -q \Phi_c - E_c)/k_B T$
contain the quasi-Fermi level information and $\tau$ is the capture time. 
The capture time represents scattering processes attributed to carrier–carrier and carrier-LO phonon 
interactions involving bound quantum well states (of which, it is generally assumed, 
carrier-LO phonon is dominating in the case considered). The net capture rate $C$ 
is added to the continuity equations as a recombination term. 

In a similar way scattering of holes is treated, with their own characteristic 
capture time. These parameters are specified in Sentaurus by the keywords $QWeScatTime$ and
$QWhScatTime$. Their default values, $8 \cdot 10^{-13} s$ and $4 \cdot 10^{-13} s$, respectively, 
correspond reasonably well to these based on theory (\cite{Hernandez}, \cite{Blom}), while
photoluminescence spectroscopy results give values of an order of $3-20 ps$ \cite{Blom}.
In most of our modelling, if not indicated otherwise, we use also default values
of electron and hole mobility, represented in Sentaurus by parameters
$eQWMobility = 9200 cm^2/Vs$ and $hQWMobility = 400 cm^2/Vs$, since we do not observe a noticable changes 
of I-V-L laser characteristics when these parameters change a few times in any direction.

For shallow quantum wells, the energy transfer during scattering can only occur in a limited
range. In the limit of elastic scattering, the net capture rate is then approximated by:

\begin{equation}\label{capture_rate_shallow}
C = \left(\frac{F_{3/2}(\eta_c)}{F_{1/2}(\eta_c)} -  \frac{F_{3/2}(\eta_b)}{F_{1/2}(\eta_b)}\right) \cdot \frac {n_c}{\tau}
\end{equation}

where $F_m$ is the Complete Fermi-Dirac integral of the order of $m$. The shallow quantum well model is
activated by the keyword $QWShallow$. 

It should be pointed out that Equations 
\ref{capture_rate_deep} and 
\ref{capture_rate_shallow}, for deep
and shallow quantum wells, respectively, while provide a convenient,
intuitive description of carriers scattering and capture on QW, these are approximate only.
In particular, there is no dependence of capture time on energy of unbound carriers there 
and no periodic oscillations as a function of QW size (\cite{Hernandez}, \cite{Blom}). 
GaAs/AlGaAs are considered to have deep quantum wells. However, as we have shown, the results
of our modelling indicate on a strong role of bound QW states located very closely to the 
offset energy levels of quantum wells. For these reasons, we did not restrict our calculations to 
deep- or shallow- QW models but used instead the full model available in Sentaurus.

Equations 
\ref{current_density} - 
\ref{electron_emission} all depend on density of bound states in QW. 
We expect hence that current through the QW will be proportional to the density of all bound states in
QW. In effective mass approximation, the two-dimensional density of electron states
within each QW subband $n$ equals (\cite{Piprek})

\begin{equation}\label{density_of_states}
\begin{array}{ll}
N_n(E) = \frac{m_n}{\pi \hbar ^2},~~~ for~ E > E_n
\end{array}
\end{equation}

Hence, the current should be proportional to the \emph{number of bound states} $\times$ \emph{carrier mass}.
The quantity computed this way (with a certain multiplication factor) 
is represented by large circles in Figure 
\ref{qw300_talk0A}. 
Though it must not be exact (for instance, no difference
in scattering rates for electrons and holes is accounted for), it fits reasonably well $I(d_a)$ dependence.

In our modelling, we assumed scattering times independent of QW width 
(modeling is performed however for a few sets of scattering times). 
In Birner and Ferreira results (\cite{Ferreira}, \cite{Birner2}), 
scattering time steeply increases with decrease of QW width, below the width of around 6 nm,
and slowly, monotonically decreases when it becomes larger that about 6 nm, with a value of around $10^{-12} s$ at 7 nm.

The default value of scattering time used by Synopsys is $8\cdot 10^{-13} s$, for electrons. That suggests that results represented by data 
on curves B and C in Figure \ref{qw300_BCDE} should be closest to these expected experimentally. At the same time,
changing of scattering times with QW width should not diminish the existence of steps since these changes are monotonic.
Also, the steps should become more pronounced at large QW widths, 
as well the current values should increase in a steeper way than would follow from Figures \ref{qw300_talk0A} or \ref{qw300_BCDE}.
It ought however to be remembered that in Birner\'s example results, only elecrons scattering is taken into account 
and only transitions between the lowest bound electron states. The results reported in literature
often predict a quasi-periodic dependence of scattering times on QW width (\cite{Hernandez}, \cite{Blom}, \cite{Mosko}).


\section{Summary and Conclusions}
\label{sec:summary} 

When performing modeling of laser characteristics as a function of the width of active region 
we noticed a non-monotonic, discontinuous dependence of $I(d_a)$ (when measured at constant voltage applied). 
A careful analysis of the data led us to the hypothesis that discontinuities occur when the most 
upper QW, bound energy states are found very close to the conduction or valence band energy offsets. 
The effect, hence, is thought to be related to changes in density of states of carriers from one hand,
and to fast changes in carrier transfer matrix through QW for QW bound states close 
to $E_{CBO}$ or $E_{VBO}$. As such, it ought to be more pronounced at lower temperatures, as 
confirmed by results of modeling $I(d_a)$ at liquid Nitrogen temperature (\cite{Matukhin}).

The effect is observed also when modelling current as a function of QW depth (Al concentration in waveguide).

Therefore, we concluded that a similar effect will be present also in modelling $I$ as a function
of temperature. In that case however a carefull design of laser properties is needed, in such a way
that a transition of the most upper QW energy state will pass through an edge of quantum well when temperature is swapt. 

We expect also that performing measurements on laser devices under uniaxial or hydrostatic pressure might
provide experimental evidence on signficance of these effects in real devices.

These observations are potentially important for proper designing of semiconducting lasers 
(choice of Al concentrations, thickness of the active region, etc), and, potentially, 
might be useful for designing a kind of quantum level spectroscopy tool for testing lasers 
for technological applications. 

Comparison of results with these obtained when ballistic transport is included is very desirable.


\bibliography{references}

\end{document}